\documentclass[journal] {IEEEtran}
\usepackage{amsfonts}
\usepackage{amsfonts}
\usepackage{amsfonts}
\usepackage{amsfonts}
\usepackage{mathrsfs}
\usepackage{cite}
\usepackage{graphicx}
\usepackage{amsmath}
\usepackage{amssymb}
\usepackage{stfloats}
\ifCLASSINFOpdf
\else
\fi
\begin{document}
\title{A Differential Feedback Scheme Exploiting the Temporal and Spectral Correlation}

\author{\IEEEauthorblockN{Mingxin~Zhou,
Leiming~Zhang,~\IEEEmembership{Member,~IEEE,}
 Lingyang~Song,~\IEEEmembership{Senior Member,~IEEE,}
  and Merouane~Debbah,~\IEEEmembership{Senior Member,~IEEE,}\\}
\thanks{Manuscript received November 25, 2012; revised February 20, 2013; accepted May 4, 2013. The associate editor coordinating the review
of this paper and approving it for publication was Prof.
Xianbing Wang. 
This work was partially supported by the National 973 project under
grant 2013CB336700, National Nature Science Foundation of China
under grant number 61222104 and 61061130561, the Ph.D. Programs
Foundation of Ministry of Education of China under grant number
20110001110102, and the Opening Project of Key Laboratory of Cognitive Radio and Information Processing (Guilin University of Electronic Technology).}
\thanks{Copyright (c) 2013 IEEE. Personal use of this material is permitted. However, permission to use this material for any other purposes must be obtained from the IEEE by sending a request to pubs-permissions@ieee.org.}
\thanks{M. Zhou, and L.
Song are with the State Key Laboratory of Advanced Optical
Communication Systems and Networks, School of Electronics
Engineering and Computer Science, Peking University, Beijing, P. R.
China, 100871. (e-mail: \{mingxin.zhou,
lingyang.song\}@pku.edu.cn).}
\thanks{L. Zhang is with Huawei Technol., Beijing, P. R. China, 100095. (e-mail: leiming.zhang@pku.edu.cn)}
\thanks{M. Debbah is with SUPELEC, Alcatel-Lucent Chair in Flexible Radio, 3 rue Joliot-Curie,
FR-91192 Gif Sur Yvette, France. (e-mail:
merouane.debbah@supelec.fr)} }


%
\maketitle
\begin{abstract}
Channel state information (CSI) provided by limited feedback channel
can be utilized to increase the system throughput. However, in
multiple input multiple output (MIMO) systems, the signaling
overhead realizing this CSI feedback can be quite large, while the
capacity of the uplink feedback channel is typically limited. Hence,
it is crucial to reduce the amount of feedback bits. Prior work on
limited feedback compression commonly adopted the block fading
channel model where only temporal or spectral correlation in
wireless channel is considered. In this paper, we propose a
differential feedback scheme with full use of the temporal and
spectral correlations to reduce the feedback load. Then, the minimal
differential feedback rate over MIMO time-frequency (or doubly)
selective fading channel is investigated. Finally, the analysis is
verified by simulation results.
\end{abstract}


\IEEEpeerreviewmaketitle

\section{Introduction}
In multiple input and multiple output (MIMO) systems, channel
adaptive techniques (e.g., water-filling, interference alignment,
beamforming, etc.) can enhance the spectral efficiency or the
capacity of the system. However, these channel adaptive techniques
require accurate channel conditions, often referred to channel state
information (CSI). Oftentimes, in a Frequency-Division Duplex (FDD)
setting, CSI is estimated at the receiver and conveyed to the
transmitter via a feedback channel. In recent years, CSI feedback
problems have been intensively studied, due to its potential
benefits to the MIMO systems \cite{Ref:1,Ref:2}. It is significant
to explore how to reduce the feedback load, due to the uplink
feedback channel limitation.

In \cite{Ref:3}, four feedback rate reduction approaches were
reviewed, where the lossy compression using the properties of the
fading process was considered best. When the wireless channel
experiences temporal-correlated fading, modeled as a finite-state
Markov chain, the amount of CSI feedback bits can be reduced by
ignoring the states occurring with small probabilities
\cite{Ref:4,Ref:5,Ref:6,Ref:7,Ref:zmx}. The feedback rate in
frequency-selective fading channels was studied in
\cite{Ref:9,Ref:w.h.}, by exploiting the frequency correlation.

In summary, all the above works mainly focus on feedback rate
compression considering either temporal correlation or spectral
correlation. However, doubly selective fading channels are more
frequently encountered in wireless communications as the desired
data rate and mobility grow simultaneously. To the best knowledge of
the authors, the scheme of making full use of the two-dimensional
correlations is not yet well studied. Using both of the orthogonal
dimensional correlations in a cooperated way, the feedback overhead
can be further reduced in the doubly selective fading channels.
Thus, in this paper, we derive the minimal feedback rate using both
the temporal and spectral correlations.

The main contributions of the present paper can be briefly
summarized as:1) We discuss
the minimal feedback rate without differential feedback. 2) We propose a differential feedback scheme by exploiting the temporal and spectral
correlations, and 3) We derive
the minimal differential feedback rate expression over MIMO doubly
selective fading channel.

The rest of the paper is organized as follows: In Section~II, we
describe the differential feedback model as well as the statistics
of the doubly selective fading channel. In Section~III, we propose a
differential feedback scheme by exploiting the two-dimensional
correlations and derive the minimal feedback rate. In Section~IV, we
provide some simulation results showing the performance of the
proposed scheme.

\section{System Model}
In this paper, we assume that the down-link channel is a mobile
wireless channel which is always correlated in time and frequency
domains, while the up-link channel is a limited feedback channel.

\subsection{Statistics of the down-link channel}
Since the channel corresponding to each antenna is independent and
with the same statistics, we can describe the separation property of
the channel frequency response $H(t,f)$ at time $t$ for an arbitrary
transmit and receive antenna pair~\cite{Ref:11}
\begin{align}\label{eq:sm_time_frequency_correlation}
{r_H}\left( {\Delta t,\Delta f} \right) &= \mathbb{E}\left\{ {H\left( {t + \Delta t,f + \Delta f} \right){H^*}\left( {t,f} \right)} \right\}\nonumber\\
 &= \sigma _H^2{r_t}\left( {\Delta t} \right){r_f}\left( {\Delta f}
 \right),
\end{align}
where $\mathbb{E} \left\{ \cdot \right\}$ denotes expectation
function, the superscript $(\cdot)^*$ denotes complex conjugate.
$\sigma _H^2$ is the power of the channel frequency response.
${r_t}\left( {\Delta t} \right)$ and ${r_f}\left( {\Delta f}
 \right)$ denotes the temporal and spectral correlation functions, respectively.

Assuming that the channel frequency response stays constant within
the symbol period $t_s$ and the subchannel spacing $f_s$, the
correlation function for different periods and subchannels is
written as
\begin{equation}\label{eq:sm_discrete_correlation_function}
{r_H}\left[ {\Delta m,\Delta n} \right] = \sigma_H^2 {r_t}\left[
\Delta m \right]{r_f}\left[\Delta n \right],
\end{equation}
where ${r_t}\left[ \Delta m \right] = {\kern 1pt} {r_t}\left(
{\Delta m{t_s}} \right)$ and ${r_f}\left[ \Delta n \right]{\kern
1pt} {\kern 1pt} = r\left( {\Delta n{f_s}} \right)$.

Furthermore, if we just consider the time domain, the correlated
channel can be modeled as a time-domain first-order autoregressive
process (AR1)\cite{Ref:4}
\begin{equation}\label{eq:sm_time_AR1}
{{H}_{m,n}} = {\alpha _t}{{H}_{m-1,n}} + \sqrt {1 - \alpha _{_t}^2}
{{W}_t},
\end{equation}
where ${{H}_{m,n}}$ denotes the channel coefficient of the $m$th
symbol interval and the $n$th subchannel, ${{W}_t}$ is a complex
white noise variable, which is independent of ${{H}_{m-1,n}}$, with
variance $\sigma_H^2$. The parameter $\alpha_t$ is the time
autocorrelation coefficient, which is given by the zero-order Bessel
function of first kind $\alpha_t={r_t}[ 1 ] = {J_0}\left( {2\pi
{f_d}t_s } \right)$, where $f_d$ is the Doppler
frequency~\cite{Ref:12}.

Similarly, if we just consider the frequency domain, the correlated
channel can also be represented as a frequency-domain AR1
\cite{Ref:9}
\begin{equation}\label{eq:sm_freq_AR1}
{{H}_{m,n}} = {\alpha _f}{{H}_{m,n-1}} + \sqrt {1 - \alpha _f^2}
{{W}_f},
\end{equation}
where ${{W}_f}$ is a complex white noise variable, which is
independent of ${{H}_{m,n-1}}$, with variance $\sigma_H^2$. The
parameter $\alpha_f$ determines the correlation between the
subchannels, which is given by $\alpha_f={r_f}\left[ 1
\right]=\frac{1}{\sqrt{1+(2\pi f_s \Delta )^2}}$, where $\Delta$ is
the root mean square delay spread~\cite{Ref:12}.

\subsection{Differential Feedback Model}
\begin{figure}[h!]
\centering
\includegraphics[width=3.3in]{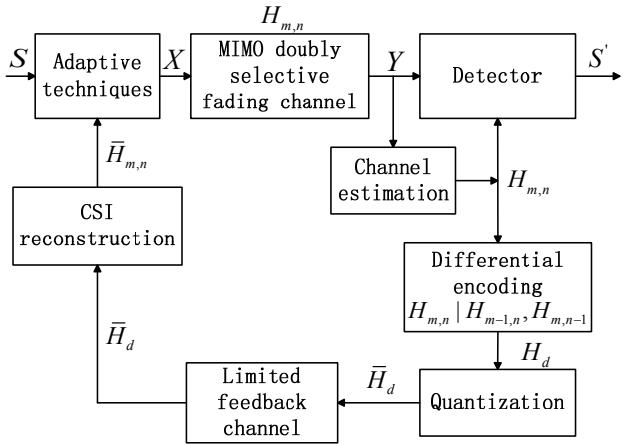}
\caption{System model of the differential feedback over MIMO doubly
selective fading channel} \label{fig:sm}
\end{figure}

The system model with differential feedback is illustrated in
Fig.~\ref{fig:sm}. By using differential feedback scheme, the
receiver just feeds back the differential CSI.

We suppose that there are $N_t$ and $N_r$ antennas at the
transmitter and receiver, respectively. The received signal vector
at the $m$th symbol interval and the $n$th subchannel is given by
\begin{equation}\label{eq:SM_received_signal}
\mathbf{y}_{m,n}=\mathbf{H}_{m,n}\mathbf{x}_{m,n}+\mathbf{n}_{m,n}.
\end{equation}
In the above expression, $\mathbf{y}_{m,n}$ denotes the $N_r \times
1$ received vector at the $m$th symbol interval and the $n$th
subchannel. $\mathbf{H}_{m,n}$, a $N_r \times N_t$ channel fading
matrix, is the frequency response of the channel. The entries are
assumed independent and identically distributed (i.i.d.), obeying a
complex Gaussian distribution with zero-mean and variance
$\sigma_H^2$. Different antennas have the same characteristic in
temporal and spectral correlations, $\alpha_t$ and $\alpha_f$,
respectively. Besides, there is no spatial correlation between
different antennas. $\mathbf{x}_{m,n}$ denotes the $N_t \times 1$
transmitter signal vector and is assumed to have unit variance.
$\mathbf{n}_{m,n}$ is a ${N_r} \times 1$ additive white Gaussian
noise (AWGN) vector with zero-mean and variance $\sigma_0^2$. Both
$\mathbf{x}_{m,n}$ and $\mathbf{n}_{m,n}$ are independent for
different $m$'s and $n$'s.

Through CSI quantization, the feedback channel output is written as
\cite{Ref:13,Ref:8,Ref:17}
\begin{equation}\label{eq:sm_quan_channel}
\mathbf{H}_{m,n} = \bar {\mathbf{H}}_{m,n} + \mathbf{E}_{m,n},
\end{equation}
where $\bar {\mathbf{H}}_{m,n}$ denotes the channel quantization
matrix, and $\mathbf{E}_{m,n}$ is the independent additive
quantization distortion matrix whose entries are zero-mean and with
variance $\frac{D}{{{N_r}{N_t}}}$, where $D$ represents the channel
quantization distortion constraint.

The differential feedback is under consideration as shown in
Fig.~\ref{fig:sm}. We can use the previous CSI to forecast the
present CSI ${\bf H}_{m,n}$ at the transmitter
\begin{equation}\label{eq:sm_channel_esti}
\hat{\bf{H}}_{m,n} = {a_1}{{\bf{H}}_{m - 1,n}} + {a_2}{{\bf{H}}_{m,n
- 1}},
\end{equation}
where $a_1$ and $a_2$ are the coefficients of the channel predictor
which will be calculated by using the minimum mean square error
(MMSE) principle in the next section. Meanwhile, the receiver
calculates the differential CSI, given the previous ones. The
differential CSI can be formulated as
\begin{equation}\label{eq:SM_diff}
\mathbf{H}_d={\rm Diff}\left(\mathbf{H}_{m,n}| {\mathbf{H}}_{m-1,n},
{\mathbf{H}}_{m,n-1}\right),
\end{equation}
where $\mathbf{H}_d$ represents the differential CSI which obviously
is the prediction error, and $Diff\left( { \cdot } \right)$ is the
differential function. Then through limited feedback channel, ${\bf
H}_d$ should be quantized and fed back.

Finally, The CSI reconstructed by combining the differential one and
the channel prediction is utilized by the channel adaptive
techniques. In this paper, we adopt the water-filling precoder,
however, the analysis and conclusions given in this paper are also
valid for other adaptive techniques.

The channel quantization matrix is decomposed as ${\bar{ \bf
H}}_{m,n}=\bf {\bar U} {\bar \Sigma} {\bar V}^+$ using singular
value decomposition (SVD) at the transmitter. $\bf {\bar U}$ and
$\bf {\bar V}$ are unitary matrixes, and $\bf {\bar \Sigma}$ is a
non-negative diagonal matrix composed of eigenvalues of ${\bar{ \bf
H}}_{m,n}$.

With the water-filling precoder, the closed-loop capacity can be
obtained as \cite{Ref:13,Ref:8,Ref:17}
\begin{equation}\label{eq:sm_erg_capacity}
{C_{erg}} = \mathbb{E}\left[ {\log \det \left( {{{\bf{I}}_{{N_r}}} +
{\bf{J}} \cdot {{\bf{J}}^ + }\left( {{{\bf{F}}^{ - 1}}} \right)}
\right)} \right],
\end{equation}
where ${\bf{J}}={\bf {\bar H}}_{m,n} \bf{\bar V} {\bar Z}$,
${\bf{J}}_e={\bf{E}}_{m,n}\bf{\bar V} {\bar Z}$, and
${\bf{F}}=\frac{1}{{{A^2}}}{{\bf{I}}_{{N_r}}} +
\mathbb{E}\left[{\bf{J}}_e{\bf{J}}_e^+ | {\bf{J}}\right]$, where $A$
represents the amplitude of signal symbol, and $\bar {\bf Z}$
denotes a diagonal matrix determined by water-filling
\cite{Ref:13,Ref:8,Ref:17}
\begin{equation}\label{eq:sm_wf}
\left\{ \begin{array}{l} \bar z_i^2 = \left\{ \begin{array}{l}
\bar \mu  - {(\bar \gamma _{i,i}^2{A^2})^{ - 1}},\\
0,
\end{array} \right.\begin{array}{*{20}{c}}
{\bar \gamma _{i,i}^2{A^2} \ge \bar\mu _{}^{ - 1}}\\
{{\rm{otherwise}}}
\end{array}\\
\sum\limits_{i = 1}^{{N_t}} {\bar z_i^2{A^2} = {N_t}{A^2}},
\;\;\;\;\;\;\;{\rm power\; constraint}
\end{array} \right.,
\end{equation}
where $\bar \gamma _{i,i}, i=1, 2, ... , N_t$ are the entries of
$\bar \Sigma$, $\bar \mu$ denotes a cut-off value chosen to meet the
power constraint.

It is obvious from (\ref{eq:sm_erg_capacity}) that the closed-loop
ergodic capacity is determined by ${\bf H}_{m,n}$ and ${\bf \bar
H}_{m,n}$, and the loss of capacity is mainly caused by the
quantization error. Therefore, given the limited feedback channel,
the capacity can be enhanced by exploiting the channel correlations
to reduce the quantization error.

\section{Minimal Differential Feedback Rate}
In this section, exploiting the temporal and spectral correlations,
we study the minimal feedback rate that denotes the minimal feedback
bits required per block to preserve the given channel quantization
distortion.

We first describe the feedback rate using normal quantization.
Without differential feedback scheme, the receiver feeds back
${{\bf{H}}_{m,n}}$ to the transmitter. The information entropy of a
Gaussian variable $X$ with variance $\sigma^2$ is represented as
\cite{Ref:14}
\begin{equation}\label{eq:MFR_info_entrpy}
h\left( X \right) = \frac{1}{2}\log 2\pi e \sigma^2.
\end{equation}
Thus, the feedback load has positive relation with $\sigma_H^2$.

Furthermore, taking quantization of the channel matrix into
consideration, the feedback rate is determined by the rate
distortion theory of continuous-amplitude sources \cite{Ref:14}
\begin{equation}\label{eq:MFR_minfb}
R = \inf \left\{ {I\left( {{{\bf{H}}_{m,n}};{{{\bf{\bar H}}}_{m,n}}}
\right):\mathbb{E} \left[ {d\left( {{{\bf{H}}_{m,n}};{{{\bf\bar
H}}_{m,n}}} \right)} \right] \le D} \right\},
\end{equation}
where $\inf\{\cdot\}$ denotes infimum function, ${I\left(
{{{\bf{H}}_{m,n}};{{{\bf{\bar H}}}_{m,n}}} \right)}$ denotes the
mutual information between ${{{{\bf{\bar H}}}_{m,n}}}$ and ${{{\bf{
H}}}_{m,n}}$, and $d\left( {{{\bf H}_{m,n}};{{{\bf{\bar H}}}_{m,n}}}
\right) = {\left\| {{{\bf H}_{m,n}} - {{{\bf{\bar H}}}_{m,n}}}
\right\|^2}$ denotes the channel quantization distortion which is
constrained by $D$.

Since the entries of $\bf{H}$ and $\bar{\bf{H}}$ are i.i.d. complex
Gaussian variables, the feedback rate can be written as
\begin{equation}\label{eq:MFR_1d_minfb}
R = \inf \left\{ {{N_t}{N_r}I\left( {{H_{m,n}};{{\bar H}_{m,n}}}
\right):\mathbb{E}[d({H_{m,n}},{{\bar H}_{m,n}})] \le d} \right\},
\end{equation}
where $d = \frac{D}{{{N_t}{N_r}}}$ is the one-dimensional average
channel quantization distortion. $H_{m,n}$ and $\bar{H}_{m,n}$
represent the entries of ${\bf{H}}_{m,n}$, $\bar{{\bf{H}}}_{m,n}$,
respectively. Also, from (\ref{eq:sm_quan_channel}) the
one-dimensional channel quantization is written as
\begin{equation}\label{eq:MFR_1d_quan}
H_{m,n} = \bar H_{m,n} + E_{m,n}.
\end{equation}
The mutual information can be written as
\begin{equation}\label{eq:MFR_mutl_info}
I\left( {{H_{m,n}};{{\bar H}_{m,n}}} \right) = h\left( {{H_{m,n}}}
\right) - h\left( {{H_{m,n}}\left| {{{\bar H}_{m,n}},} \right.}
\right).
\end{equation}
Combining (\ref{eq:MFR_1d_quan}), (\ref{eq:MFR_mutl_info}) can be
rewritten as
\begin{equation}\label{eq:MFR_mutl_info1}
I\left( {{H_{m,n}};{{\bar H}_{m,n}}} \right) \ge h\left( {{H_{m,n}}}
\right) - h\left( E_{m,n}\right).
\end{equation}
Substituting (\ref{eq:MFR_info_entrpy}) and
(\ref{eq:MFR_mutl_info1}) into (\ref{eq:MFR_1d_minfb}), we obtain

\begin{equation}\label{eq:MFR_no_finalrate}
R = N_rN_t \log \left( {\frac{\sigma _H^2}{d}} \right).
\end{equation}

From (\ref{eq:MFR_no_finalrate}), the feedback rate required for the
non-differential feedback is very large. Nevertheless, by employing
the temporal and spectral correlations, we can use the differential
feedback scheme to reduce the feedback bits significantly. The
transmitter can predict the present CSI ${\mathbf{H}}_{m,n}$
depending on the previous ones in time domain ${\mathbf{H}}_{m-1,n}$
and frequency domain ${\mathbf{H}}_{m-1,n}$. Then, the receiver
quantizes $\mathbf{H}_d$ ,or equivalently, the error of the channel
prediction, and feeds back to the transmitter. Finally, the
transmitter reconstructs the CSI by both the channel prediction and
the differential CSI. It is obvious that the more accurate the
channel is predicted, the less bits is fed back from the receiver.
As $ {\mathbf{H}}_{m-1,n}, {\mathbf{H}}_{m,n-1}$ and
$\mathbf{H}_{m,n}$ are correlated, an MMSE channel predictor can be
constructed as (\ref{eq:sm_channel_esti}), where the coefficients
$a_1$ and $a_2$ are selected to minimize
\begin{equation}\label{eq:MFR_esti_MSE}
{\rm{MSE}}\left( {{a_1},{a_2}} \right) = \mathbb{E}{\left|
{{{{\bf{\hat H}}}_{m,n}} - {{\bf{H}}_{m,n}}} \right|^2}.
\end{equation}
The MSE represents the statistical difference between the predicted
value and the true one. We can obtain the minimized quantization
bits by minimizing the MSE.

We can rewrite ${\bf{H}}_{m,n}$ as
\begin{equation}\label{eq:MFR_Hmn_rewrite}
{\bf{H}}_{m,n} = {{{\bf{\hat H}}}_{m,n}} + {{\bf{H}}_d} =
{a_1}{{\bf{H}}_{m - 1,n}} + {a_2}{{\bf{H}}_{m,n - 1}} +
{{\bf{H}}_d},
\end{equation}
where ${{\bf{H}}_d}$ is the differential feedback load to minimize.
By the orthogonality principle \cite{Ref:15}, $a_1,a_2$ are
determined by
\begin{equation}\label{eq:MFR_orth_principle}
\left\{ \begin{array}{l}
\mathbb{E}\left[ {\left( {{{\bf{H}}_{m,n}} - {a_1}{{\bf{H}}_{m - 1,n}} - {a_2}{{\bf{H}}_{m,n - 1}}} \right){{\bf{H}}_{m - 1,n}}} \right] = 0\\
\mathbb{E}\left[ {\left( {{{\bf{H}}_{m,n}} - {a_1}{{\bf{H}}_{m -
1,n}} - {a_2}{{\bf{H}}_{m,n - 1}}} \right){{\bf{H}}_{m,n - 1}}}
\right] = 0
\end{array} \right..
\end{equation}

Since the entries of
${{\bf{H}}_{m,n}}$,${{\bf{H}}_{m-1,n}}$,${{\bf{H}}_{m,n-1}}$ are
i.i.d. complex Gaussian variables, the orthogonality principle can
be rewritten as
\begin{equation}\label{eq:MFR_orth_principle_single}
\left\{ \begin{array}{l}
\mathbb{E}\left[ {\left( {{H_{m,n}} - {a_1}{H_{m - 1,n}} - {a_2}{H_{m,n - 1}}} \right){H_{m - 1,n}}} \right] = 0\\
\mathbb{E}\left[ {\left( {{H_{m,n}} - {a_1}{H_{m - 1,n}} -
{a_2}{H_{m,n - 1}}} \right){H_{m,n - 1}}} \right] = 0
\end{array} \right..
\end{equation}
Moreover, the one-dimensional frequency response of the channel can
be represented as
\begin{equation}\label{eq:MFR_1d_Hmn}
{{H}}_{m,n} = {{{{\hat H}}}_{m,n}} + {{{H}}_d} = {a_1}{{{H}}_{m -
1,n}} + {a_2}{{{H}}_{m,n - 1}} + {{{H}}_d},
\end{equation}
where $H_{m,n}$, ${{{{\hat H}}}_{m,n}}$, ${H}_{m-1,n}$, $H_{m,n-1}$
and ${{{H}}_d}$ represent the corresponding entries.

Direct calculation shows that (\ref{eq:MFR_orth_principle_single})
is equivalent to
\begin{equation}\label{eq:MFR_orth_prin_direct1}
\left\{ \begin{array}{l}
{r_H}\left[ {1,0} \right] - {a_1}{r_H}\left[ {0,0} \right] - {a_2}{r_H}\left[ {1,1} \right] = 0\\
{r_H}\left[ {0,1} \right] - {a_1}{r_H}\left[ {1,1} \right] -
{a_2}{r_H}\left[ {0,0} \right] = 0
\end{array} \right..
\end{equation}
With the separation property of the correlations of the channel
frequency response (\ref{eq:sm_discrete_correlation_function}), and
combining ${r_t}\left[ 0 \right]={r_f}\left[ 0 \right]=1$ and
$r_t[1]=\alpha_t$, $r_f[1]=\alpha_f$,
(\ref{eq:MFR_orth_prin_direct1}) can be simplified by
\begin{equation}\label{eq:MFR_orth_prin_substitue}
\left\{ \begin{array}{l}
{a_1}\sigma _H^2 + {a_2}{\alpha _t}{\alpha _f}\sigma _H^2 - {\alpha _t}\sigma _H^2 = 0\\
{a_1}{\alpha _t}{\alpha _f}\sigma _H^2 + {a_2}\sigma _H^2 - {\alpha
_f}\sigma _H^2 = 0
\end{array} \right..
\end{equation}
From (\ref{eq:MFR_orth_prin_substitue}), $a_1, a_2$ are given by
\begin{equation}\label{eq:MFR_a1_a2}
\left\{ \begin{array}{l}
{a_1} = \frac{{{\alpha _t}\left( {1 - \alpha _f^2} \right)}}{{1 - \alpha _t^2\alpha _f^2}}\\
{a_2} = \frac{{{\alpha _f}\left( {1 - \alpha _t^2} \right)}}{{1 -
\alpha _t^2\alpha _f^2}}
\end{array} \right..
\end{equation}
Combing (\ref{eq:MFR_a1_a2}) and (\ref{eq:MFR_1d_Hmn}), the
one-dimensional MSE of the channel estimator is
\begin{equation}\label{eq:MFR_aver_power_fb}
{\rm{MSE}} = {{\rm Var}}\left(H_d\right)= \sigma _H^2\left( {1 -
a_{_1}^2 - a_2^2 - 2 {a_1}{a_2}{\alpha _t}{\alpha _f}} \right).
\end{equation}
Finally, the channel estimator $\hat{\bf{H}}_{m,n}$ is given by
\begin{equation}\label{eq:MFR_chan_est_final}
\hat{\bf{H}}_{m,n}= \frac{{{\alpha _t}\left( {1 - \alpha _f^2}
\right)}}{{1 - \alpha _t^2\alpha _f^2}}{{\bf{H}}_{m - 1,n}} +
\frac{{{\alpha _f}\left( {1 - \alpha _t^2} \right)}}{{1 - \alpha
_t^2\alpha _f^2}}{{\bf{H}}_{m,n - 1}}.
\end{equation}
And combining (\ref{eq:MFR_Hmn_rewrite}) and
(\ref{eq:MFR_chan_est_final}), ${\bf{H}}_{m,n}$ is given by
\begin{equation}\label{eq:MFR_Hmn_fianl_expression}
{{\bf{H}}_{m,n}}= \frac{{{\alpha _t}\left( {1 - \alpha _f^2}
\right)}}{{1 - \alpha _t^2\alpha _f^2}}{{\bf{H}}_{m - 1,n}} +
\frac{{{\alpha _f}\left( {1 - \alpha _t^2} \right)}}{{1 - \alpha
_t^2\alpha _f^2}}{{\bf{H}}_{m,n - 1}} + {{\bf{H}}_d}.
\end{equation}

Then, through the feedback channel, the error of the channel
predictor ${\bf H}_d$ can be fed back from the transmitter to the
receiver. Similarly, from (\ref{eq:MFR_info_entrpy}), the feedback
load is positive related with ${\rm Var}(H_d)=\sigma _H^2\big( 1 -
a_{_1}^2 - a_2^2 -2{a_1}{a_2}{\alpha _t}{\alpha _f} \big)$. Because
$\frac{{\partial \rm MSE}}{{\partial {\alpha _t}}} < 0$,
$\frac{{\partial \rm MSE}}{{\partial {\alpha _f}}} < 0$, the
feedback load can be much smaller than $\sigma_H^2$, the
non-differential one, especially when the channel is highly
correlated. For example, given $\alpha_t > 0.75$, $\alpha_f
> 0.75$, then ${\left. {{\rm{MSE}}} \right|_{{\alpha _t}
> 0.75,\;{\alpha _f} > 0.75}} < {\left. {\rm MSE} \right|_{{\alpha _t} = 0.75,\;{\alpha
_f} = 0.75}} = 0.28\sigma_H^2$.

From (\ref{eq:MFR_Hmn_fianl_expression}), taking quantization impact
into consideration, the minimal differential feedback rate over
doubly selective fading channels can be calculated by the rate
distortion theory of continuous-amplitude sources in a similar way.
\begin{equation}\label{eq:MFR_minfb_tf}
R = {N_r}{N_t}\log \left\{ {a_1^2 + a_2^2 +
\frac{{2{a_1}{a_2}{\alpha _t}{\alpha _f}d}}{{\sigma _H^2}} +
\frac{{Var\left( {{H_d}} \right)}}{d}} \right\},
\end{equation}
where the channel predictor coefficients $a_1,a_2$ are determined by
${a_1} = \frac{{{\alpha _t}\left( {1 - \alpha _f^2} \right)}}{{1 -
\alpha _t^2\alpha _f^2}}$ and ${a_2} = \frac{{{\alpha _f}\left( {1 -
\alpha _t^2} \right)}}{{1 - \alpha _t^2\alpha _f^2}}$. The average
power of $H_d$ is ${{Var}}\left(H_d\right)= \sigma _H^2\left( {1 -
a_{_1}^2 - a_2^2 - 2{a_1}{a_2}{\alpha _t}{\alpha _f}} \right)$. The
detailed derivation is given in Appendix A.

The above expression gives the minimal differential feedback rate
simultaneously utilizing the temporal and spectral correlations.
From (\ref{eq:MFR_minfb_tf}), the minimal differential feedback rate
is a function of $\alpha_t, \alpha_f$ and the channel quantization
distortion $d$, and much smaller than that of the non-differential
one (\ref{eq:MFR_no_finalrate}).

\section{Simulation Results and Discussion}
In this section, we first provide the relationship between the MSE
of the predictor and the two-dimensional correlations in
Fig.~\ref{fig:MSE}. The minimal differential feedback rate over MIMO
doubly selective fading channels is given in
Fig.~\ref{fig:ft_mini_fb}. Then, a longitudinal section of
Fig.~\ref{fig:ft_mini_fb} is presented, where we assume the temporal
correlation and spectral correlation is equal. Finally, we verify
our theoretical results by a practical differential feedback system
with water-filling precoder and Lloyd's quantization algorithm
\cite{Ref:18}.

\subsection{MSE of the predictor and Minimal Differential Feedback Rate}

\begin{figure}[h!]
\centering
\includegraphics[width=3.3in]{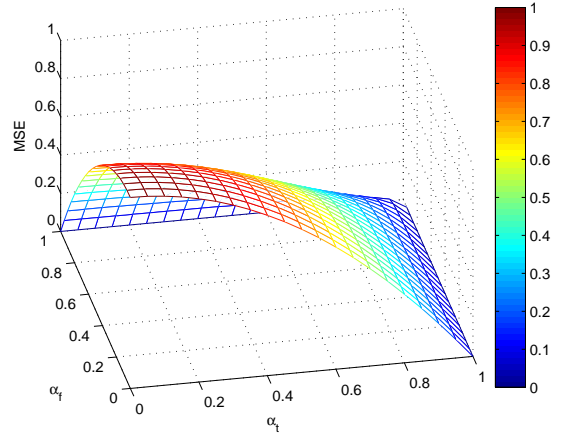}
\caption{The MSE of the predictor at the transmiter, for
$N_r=2,N_t=2,\sigma_H^2=1$ and $D=0.1$.} \label{fig:MSE}
\end{figure}

For simplicity and without loss of generality, we consider
$N_r=N_t=2$, and $\sigma_H^2=1$. Fig.~\ref{fig:MSE} presents the MSE
between the predicted value and the true value. As the temporal or
spectral correlation increases, the MSE decreases. Furthermore, when
either $\alpha_t$ or $\alpha_f$ comes to one, the MSE tends to zero.

\begin{figure}[h!]
\centering
\includegraphics[width=3.3in]{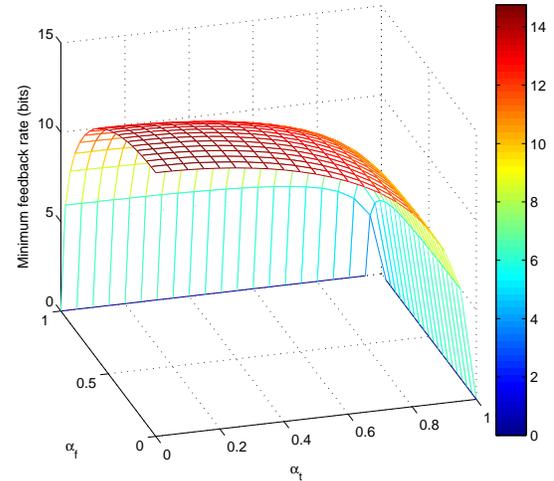}
\caption{The minimal differential feedback rate, for
$N_r=2,N_t=2,\sigma_H^2=1$ and $D=0.1$.} \label{fig:ft_mini_fb}
\end{figure}

Fig.~\ref{fig:ft_mini_fb} plots the relationship between the minimal
differential feedback rate and the two-dimensional correlations with
the channel quantization distortion $D=0.1$. It is very similar to
the MSE shown in Fig.~\ref{fig:MSE}, because it presents the minimal
bits required to quantize the differential CSI.

\begin{figure}[h!]
\centering
\includegraphics[width=3.3in]{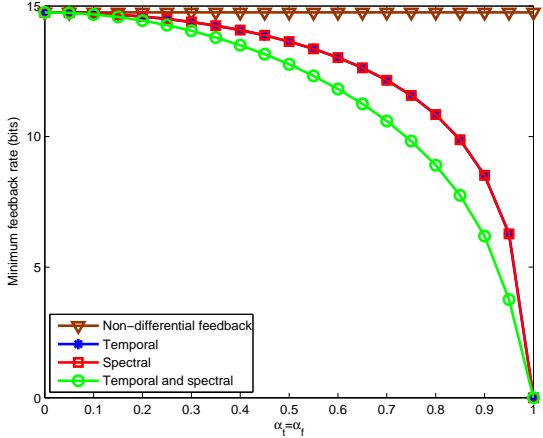}
\caption{The relationship between the minimal feedback rate and
temporal and spectral correlations, when they are equal, for
$N_r=2,N_t=2,\sigma_H^2=1$ and $D=0.1$.} \label{fig:sect_mini_fb}
\end{figure}

Additionally, because $\alpha_t$ and $\alpha_f$ could be any value,
we provide one of the longitudinal section of
Fig.~\ref{fig:ft_mini_fb} where the temporal correlation is equal to
the spectral correlation in Fig.~\ref{fig:sect_mini_fb}. For
comparison, the differential feedback compression only using
one-dimensional correlation and the non-differential feedback scheme
are also included in Fig.~\ref{fig:sect_mini_fb}. It is observed
from Fig.~\ref{fig:sect_mini_fb} that the scheme using both temporal
and spectral correlations is always better than the scheme using
only one-dimensional correlation. As the correlations increase, the
two-dimensional differential feedback compression exhibits a
significant improvement compared to one-dimensional one. This
performance advantage even reaches up to $67\%$ with
$\alpha_t=\alpha_f=0.95$.

\subsection{Differential Feedback System with Lloyd's Algorithm}
In this subsection, we consider the temporal correlation
$\alpha_t=0.9$, with carrier frequency $2$~GHz, the normalized
Doppler shift $f_d=100$~Hz and spectral correlation $\alpha_f=0.9$,
with $\Delta=8\rm\mu\rm{s}$, which is a reasonable
assumption~\cite{Ref:12}. we design a differential feedback system
using Lloyd's quantization algorithm to verify our theoretical
results \cite{Ref:18}. We use ${\rm Diff}\left(
{{{\bf{H}}_{m,n}}|{{\bf{H}}_{m - 1,n}}{{\bf{H}}_{m,n
-1}}}\right)={{\bf{H}}_{m,n}}-a_1{{\bf{H}}_{m-1,n}}-a_2{{\bf{H}}_{m,n-1}}$
as a differential function, where ${a_1} = \frac{{{\alpha _t}\left(
{1 - \alpha _f^2} \right)}}{{1 - \alpha _t^2\alpha _f^2}}$, ${a_2} =
\frac{{{\alpha _f}\left( {1 - \alpha _t^2} \right)}}{{1 - \alpha
_t^2\alpha _f^2}}$ in the two-dimensional differential feedback
compression and $a_1=\alpha_t$, $a_2=0$ in the one-dimensional one.

The feedback steps can be summarized as follows. Firstly, based on
Lloyd's quantization algorithm, the channel codebook can be
generated according to the statistics of the corresponding
differential feedback load at both transmitter and receiver.
Secondly, the receiver calculates the current differential CSI ${\bf
H}_d$. Thirdly, the differential CSI is quantized to the optimal
coodbook value ${\bf\bar H}_d$ according to the Euclidean distance.
Finally, the transmitter reconstructs the channel quantization
matrix by
${{\bf{H}}_{m,n}}=a_1{{\bf{H}}_{m-1,n}}+a_2{{\bf{H}}_{m,n-1}}+
{\bf\bar H}_d$.

\begin{figure}[h!]
\centering
\includegraphics[width=3.3in]{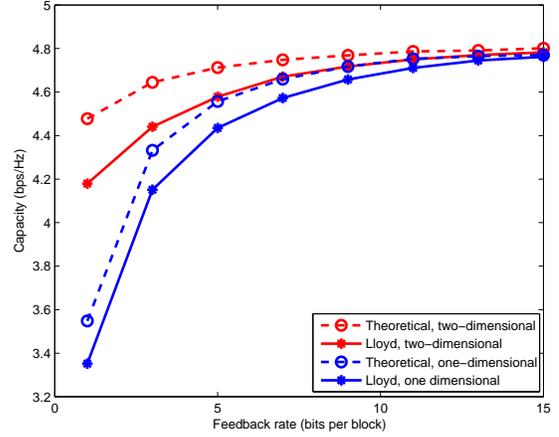}
\caption{The relationship between the ergodic capacity and feedback
rate with Lloyd's algorithm in AR1 model for
$N_r=2,N_t=2,\sigma_H^2=1$ and SNR $=5$dB.} \label{fig:cap_t_f}
\end{figure}

In Fig.~\ref{fig:cap_t_f}, we give the simulation results of the
ergodic capacity employing Lloyd's algorithm. The theoretical
capacity results are also provided in Fig.~\ref{fig:cap_t_f}. We can
see from Fig.\ref{fig:cap_t_f} that the performance of the
two-dimensional one are always better than the one-dimensional one,
which verifies our theoretical analysis.

As shown in Fig.\ref{fig:cap_t_f}, with the increase of feedback
rate $b$, the ergodic capacities increase rapidly when $b$ is small,
and then slow down in the large $b$ region, because when $b$ is
large enough, the quantization errors tend to zero. Also, the
capacities of Lloyd's quantization are lower than the theoretical
ones. The reasons are as follows. The Lloyd's algorithm is optimal
only in the sense of minimizing a variable's quantization error, but
not in data sequence compression while the channel coefficient $\bf
H$ is correlated in both temporal and spectral domain. However, the
imperfection reduces as $b$ increases, because the quantization
errors of both Lloyd's algorithm and theoretical results tend to
zero with sufficient feedback bits $b$.

\section{Conclusions}
In this paper, we have designed a differential feedback scheme
making full use of both the temporal and spectral correlation and
compared the performance with the scheme without differential
feedback. We have derived the minimal differential feedback rate for
our proposed scheme. The feedback rate to preserve the given channel
quantization distortion is significantly small compared to
non-differential one, as the channel is highly correlated in both
temporal and spectral domain. Finally, we provide simulations to
verify our analysis.

\section*{Appendix A\\ Derivation of the Minimal Differential Feedback Rate Using Temporal and Spectral Correlations}
The minimal differential feedback rate over MIMO doubly selective
fading channel can also be derived by the rate distortion theory.
Given $\bar {\bf{H}}_{m-1,n}$ and $\bar{\bf H}_{m,n-1}$ at the
transmitter, the differential feedback rate can be represented as
\begin{equation}\label{eq:apdixB_r_inf}
R \hspace{-0.9mm}=\hspace{-0.7mm} \inf
\hspace{-0.7mm}\left\{\hspace{-0.7mm} {I\hspace{-0.9mm}\left( {{{\bf
H}_{\hspace{-0.2mm}m\hspace{-0.2mm},\hspace{-0.2mm}n}}\hspace{-0.2mm};\hspace{-0.7mm}{{{\bf{\bar
H}}}_{\hspace{-0.2mm}m\hspace{-0.2mm},\hspace{-0.2mm}n}}\hspace{-0.3mm}|\hspace{-0.4mm}{{{\bf{\bar
H}}}_{\hspace{-0.2mm}m\hspace{-0.3mm} -\hspace{-0.3mm}
1\hspace{-0.2mm},\hspace{-0.2mm}n}}\hspace{-0.2mm},\hspace{-0.4mm}{{{\bf{\bar
H}}}_{\hspace{-0.2mm}m\hspace{-0.2mm},\hspace{-0.2mm}n\hspace{-0.2mm}
-\hspace{-0.2mm} 1\hspace{-0.2mm}}}}\hspace{-0.4mm} \right)}
\hspace{-0.5mm}{:\hspace{-0.9mm}\mathbb{E}\hspace{-1mm} \left[
{\hspace{-0.2mm}d\hspace{-0.7mm}\left(\hspace{-0.2mm} {{{{\bf
H}}_{\hspace{-0.2mm}m\hspace{-0.2mm},\hspace{-0.2mm}n\hspace{-0.2mm}}};\hspace{-0.6mm}{{{\bf{\bar
H}}}_{\hspace{-0.2mm}m\hspace{-0.2mm},\hspace{-0.2mm}n\hspace{-0.2mm}}}}
\hspace{-0.2mm}\right)} \hspace{-0.2mm}\right]\hspace{-1.2mm}
\le\hspace{-1.2mm} D} \hspace{-0.6mm}\right\}\hspace{-0.8mm}.
\end{equation}
Since the entries are i.i.d. complex Gaussian variables,
(\ref{eq:apdixB_r_inf}) can be written as
\begin{equation}\label{eq:apdixB_1d_r_inf}
R \hspace{-0.9mm}=\hspace{-0.7mm} \inf
\hspace{-0.7mm}\left\{\hspace{-0.7mm} {I\hspace{-0.9mm}\left( {{{
H}_{\hspace{-0.2mm}m\hspace{-0.2mm},\hspace{-0.2mm}n}}\hspace{-0.2mm};\hspace{-0.7mm}{{{{\bar
H}}}_{\hspace{-0.2mm}m\hspace{-0.2mm},\hspace{-0.2mm}n}}\hspace{-0.3mm}|\hspace{-0.4mm}{{{{\bar
H}}}_{\hspace{-0.2mm}m\hspace{-0.3mm} -\hspace{-0.3mm}
1\hspace{-0.2mm},\hspace{-0.2mm}n}}\hspace{-0.2mm},\hspace{-0.4mm}{{{{\bar
H}}}_{\hspace{-0.2mm}m\hspace{-0.2mm},\hspace{-0.2mm}n\hspace{-0.2mm}
-\hspace{-0.2mm} 1\hspace{-0.2mm}}}}\hspace{-0.4mm} \right)}
\hspace{-0.5mm}{:\hspace{-0.9mm}\mathbb{E}\hspace{-1mm} \left[
{\hspace{-0.2mm}d\hspace{-0.7mm}\left(\hspace{-0.2mm} {{{{
H}}_{\hspace{-0.2mm}m\hspace{-0.2mm},\hspace{-0.2mm}n\hspace{-0.2mm}}};\hspace{-0.6mm}{{{{\bar
H}}}_{\hspace{-0.2mm}m\hspace{-0.2mm},\hspace{-0.2mm}n\hspace{-0.2mm}}}}
\hspace{-0.2mm}\right)} \hspace{-0.2mm}\right]\hspace{-1.2mm}
\le\hspace{-1.2mm} D} \hspace{-0.6mm}\right\}\hspace{-0.8mm}.
\end{equation}
The one-dimensional channel quantization equality can be written as
\begin{align}\label{eq:apdixB_1d_quan}
&H_{m-1,n}=\bar{H}_{m-1,n}+ E_{m-1,n}\nonumber\\
&H_{m,n-1}=\bar{H}_{m,n-1}+ E_{m,n-1}.
\end{align}
Similarly, (\ref{eq:MFR_Hmn_fianl_expression}) yields
\begin{equation}\label{eq:apdixB_1d_Hmn}
{H_{m,n}} = {a_1}{H_{m - 1,n}} + {a_2}{H_{m,n - 1}} + {H_d},
\end{equation}
where ${a_1} = \frac{{{\alpha _t}\left( {1 - \alpha _f^2}
\right)}}{{1 - \alpha _t^2\alpha _f^2}}$, ${a_2} = \frac{{{\alpha
_f}\left( {1 - \alpha _t^2} \right)}}{{1 - \alpha _t^2\alpha
_f^2}}$. The conditional mutual information
$I\hspace{-0.9mm}\left(\hspace{-0.6mm}
{{H_{m,n}};\hspace{-0.5mm}{{\bar
H}_{m,n}}\hspace{-0.2mm}|\hspace{-0.2mm} {{{\bar
H}_{m\hspace{-0.1mm} -\hspace{-0.2mm}
1,n}}\hspace{-0.1mm},\hspace{-0.2mm}{{\bar H}_{m,n\hspace{-0.1mm}
-\hspace{-0.2mm} 1}}} }\hspace{-0.3mm} \right)\hspace{-0.9mm}$ can
be written as
\begin{align}\label{eq:apdixB_1d_I}
I\hspace{-0.9mm}\left(\hspace{-0.6mm}
{{H_{m,n}};\hspace{-0.5mm}{{\bar
H}_{m,n}}\hspace{-0.2mm}|\hspace{-0.2mm} {{{\bar
H}_{m\hspace{-0.1mm} -\hspace{-0.2mm}
1,n}}\hspace{-0.1mm},\hspace{-0.2mm}{{\bar H}_{m,n\hspace{-0.1mm}
-\hspace{-0.2mm} 1}}} }\hspace{-0.3mm}
\right)\hspace{-0.9mm}=\hspace{-0.8mm}
h\hspace{-0.5mm}\left(\hspace{-0.5mm}
{{H_{m,n}}\hspace{-0.1mm}|\hspace{-0.1mm} {{{\bar H}_{m
\hspace{-0.1mm}- \hspace{-0.2mm}1,n}},\hspace{-0.3mm}{{\bar H}_{m,n
\hspace{-0.1mm}- \hspace{-0.2mm}1}}} }\hspace{-0.2mm} \right)&
\\\nonumber- \hspace{-0.4mm}
h\hspace{-0.5mm}\left(\hspace{-0.5mm}
{{H_{m,n}}\hspace{-0.1mm}|\hspace{-0.1mm}{\bar H_{m,n}}, {{{\bar
H}_{m \hspace{-0.1mm}- \hspace{-0.2mm}1,n}},\hspace{-0.3mm}{{\bar
H}_{m,n \hspace{-0.1mm}- \hspace{-0.2mm}1}}} }\hspace{-0.2mm}
\right)&.
\end{align}
First, we calculate $h\left( {{H_{m,n}}\left| {{{\bar H}_{m -
1,n}},{{\bar H}_{m,n - 1}}} \right.} \right)$. Substituting
(\ref{eq:apdixB_1d_quan}) into (\ref{eq:apdixB_1d_Hmn}), it yields
that
\begin{equation}\label{eq:apdixB_1d_Hmnquan}
{H_{m,n}}\hspace{-0.8mm} = \hspace{-0.8mm}{a_1}\hspace{-0.5mm}\left(
{{{\bar H}_{m - 1,n}} \hspace{-0.5mm}+\hspace{-0.5mm} {E_{m - 1,n}}}
\right) + {a_2}\hspace{-0.8mm}\left( {{{\bar H}_{m,n -
1}}\hspace{-0.5mm} + \hspace{-0.5mm}{E_{m,n - 1}}} \right) + {H_d}.
\end{equation}
Substituting (\ref{eq:apdixB_1d_Hmnquan}) into
(\ref{eq:apdixB_1d_I}), we obtain
\begin{equation}\label{eq:apdixB_1d_I_cal1}
I \hspace{-1.1mm}=\hspace{-0.8mm}
h\hspace{-0.5mm}\left(\hspace{-0.3mm}
{{a_1}\hspace{-0.3mm}{E_{m\hspace{-0.3mm} -\hspace{-0.3mm}
1,n}}\hspace{-0.8mm} +\hspace{-0.6mm} {a_2}{E_{m,n\hspace{-0.3mm} -
\hspace{-0.3mm}1}}\hspace{-0.9mm} +\hspace{-0.6mm} {H_d}} \right)
\hspace{-0.3mm} -\hspace{-0.3mm} h\hspace{-0.5mm}\left( {{E_{m,n}}|
{{{\bar H}_{m\hspace{-0.3mm} -\hspace{-0.3mm}
1,n}},\hspace{-0.5mm}{{\bar H}_{m,n\hspace{-0.3mm} -\hspace{-0.3mm}
1}}} } \hspace{-0.3mm}\right)\hspace{-0.3mm}.
\end{equation}
Considering inequality $h\hspace{-0.5mm}\left( {{E_{m,n}}| {{{\bar
H}_{m\hspace{-0.3mm} -\hspace{-0.3mm} 1,n}},\hspace{-0.5mm}{{\bar
H}_{m,n\hspace{-0.3mm} -\hspace{-0.3mm} 1}}} }
\hspace{-0.3mm}\right)\hspace{-0.3mm}  \le\hspace{-0.3mm}
h\hspace{-0.3mm} \left( {{E_{m,n}}} \right)$
(\ref{eq:apdixB_1d_I_cal1}) can be written as
\begin{equation}\label{eq:apdixB_1d_I_cal2}
I \ge h\left( {{a_1}{E_{m - 1,n}} + {a_2}{E_{m,n - 1}} + {H_d}}
\right) - h\left( {{E_{m,n}}} \right).
\end{equation}

Since $E_{m-1,n}$, $E_{m,n-1}$ and $H_d$ are complex Gaussian
variables, and the information entropy of a Gaussian variables with
variance $\sigma^2$ is $h\left( X \right) = \frac{1}{2}\log 2\pi e
\sigma^2$, we calculate the variance of $\big( {a_1}{E_{m - 1,n}} +
{a_2}{E_{m,n - 1}} + {H_d} \big)$
\begin{align}\label{eq:apdixB_var}
Var\left( {{a_1}{E_{m - 1,n}} + {a_2}{E_{m,n - 1}} + {H_d}} \right)
= a_1^2d + a_2^2d&\\\nonumber + Var
\left({{H_d}}^2\right)+2{a_1}{a_2}r\left( E_{m-1,n},
E_{m,n-1}\right)&.
\end{align}
Now we give the derivation of the correlation function of two noise
terms $r\left( E_{m-1,n}, E_{m,n-1}\right)$. From
(\ref{eq:apdixB_1d_quan}), the quantization error can be decomposed
into two parts
\begin{equation}\label{eq:apdixB_quan_rewrt}
\begin{array}{l}
{E_{m - 1,n}} = \frac{{\sigma _H^2 - \sigma _{\bar H}^2}}{{\sigma _H^2}}{H_{m - 1,n}} + {\psi _{m - 1,n}}\\
{E_{m,n - 1}} = \frac{{\sigma _H^2 - \sigma _{\bar H}^2}}{{\sigma
_H^2}}{H_{m,n - 1}} + {\psi _{m,n - 1}}
\end{array},
\end{equation}
where
\begin{equation}\label{eq:apdixB_quan_rewrt_n}
\begin{array}{l}
{\psi _{m,n - 1}} = {{\bar H}_{m,n - 1}} - \frac{{\sigma _{\bar H}^2}}{{\sigma _H^2}}{H_{m,n - 1}}\\
{\psi _{m - 1,n}} = {{\bar H}_{m - 1,n}} - \frac{{\sigma _{\bar
H}^2}}{{\sigma _H^2}}{H_{m - 1,n}}
\end{array},
\end{equation}
$\psi$ is a Gaussian variable with zero-mean and variance
$\frac{{\sigma _{\bar H}^2\left( {\sigma _H^2 - \sigma _{\bar H}^2}
\right)}}{{\sigma _H^2}}$, independent with $H$.

Then the correlation function of $E_{m-1,n}$ and $E_{m,n-1}$ can be
calculated as
\begin{equation}\label{eq:apdixB_nois_corr}
r\left( {{E_{m - 1,n}},{E_{m,n - 1}}} \right) = \frac{{{{\left(
{\sigma _H^2 - \sigma _{\bar H}^2} \right)}^2}}}{{\sigma
_H^2}}{\alpha _t}{\alpha _f} = \frac{d^2}{{\sigma _H^2}}{\alpha
_t}{\alpha _f}.
\end{equation}
Substituting (\ref{eq:apdixB_nois_corr}) into (\ref{eq:apdixB_var}),
we obtain
\begin{align}\label{eq:apdixB_var_cal}
Var\left( {{a_1}{E_{m - 1,n}} + {a_2}{E_{m,n - 1}} + {H_d}} \right)
= a_1^2d+ a_2^2d  &\\\nonumber+ \sigma
_{{H_d}}^2+2{a_1}{a_2}\frac{{{d^2}}}{{\sigma _H^2}}{\alpha
_t}{\alpha _f}&.
\end{align}

From (\ref{eq:apdixB_1d_r_inf}), (\ref{eq:apdixB_1d_I_cal2}) and
(\ref{eq:apdixB_var_cal}), it yields that
\begin{equation}\label{eq:apdixB_final}
R = {N_r}{N_t}\log \left\{ {a_1^2 + a_2^2 +
\frac{{2{a_1}{a_2}{\alpha _t}{\alpha _f}d}}{{\sigma _H^2}} +
\frac{{Var\left( {{H_d}} \right)}}{d}} \right\},
\end{equation}
where ${a_1} = \frac{{{\alpha _t}\left( {1 - \alpha _f^2}
\right)}}{{1 - \alpha _t^2\alpha _f^2}}$, ${a_2} = \frac{{{\alpha
_f}\left( {1 - \alpha _t^2} \right)}}{{1 - \alpha _t^2\alpha _f^2}}$
and $Var\left( {{H_d}} \right) = \sigma _H^2\left( {1 - a_{_1}^2 -
a_2^2 - 2{a_1}{a_2}{\alpha _t}{\alpha _f}} \right)$.


\end{document}